\documentclass[10pt,a4paper,twocolumn,english,amssymb,nofootinbib,superscriptaddress,showpacs]{revtex4}
\usepackage{lmodern}

\usepackage[T1]{fontenc}
\usepackage[latin9]{inputenc}
\usepackage{color}
\usepackage{babel}
\usepackage{amsmath}
\usepackage{amssymb}
\usepackage{esint}
\usepackage[unicode=true,pdfusetitle,
 bookmarks=true,bookmarksnumbered=false,bookmarksopen=false,
 breaklinks=false,pdfborder={0 0 1},backref=false,colorlinks=false]
 {hyperref}

\makeatletter

\pdfpageheight\paperheight
\pdfpagewidth\paperwidth

\@ifundefined{textcolor}{}
{%
 \definecolor{BLACK}{gray}{0}
 \definecolor{WHITE}{gray}{1}
 \definecolor{RED}{rgb}{1,0,0}
 \definecolor{GREEN}{rgb}{0,1,0}
 \definecolor{BLUE}{rgb}{0,0,1}
 \definecolor{CYAN}{cmyk}{1,0,0,0}
 \definecolor{MAGENTA}{cmyk}{0,1,0,0}
 \definecolor{YELLOW}{cmyk}{0,0,1,0}
}

\usepackage{babel}
\@ifundefined{definecolor}{\usepackage{color}}{}
\@ifundefined{definecolor}{\usepackage{color}}{}
\usepackage{babel}

\@ifundefined{definecolor}{\usepackage{color}}{}
\usepackage{babel}

\pdfpageheight\paperheight
\pdfpagewidth\paperwidth

\@ifundefined{textcolor}{}{%
 \definecolor{BLACK}{gray}{0}
 \definecolor{WHITE}{gray}{1}
 \definecolor{RED}{rgb}{1,0,0}
 \definecolor{GREEN}{rgb}{0,1,0}
 \definecolor{BLUE}{rgb}{0,0,1}
 \definecolor{CYAN}{cmyk}{1,0,0,0}
 \definecolor{MAGENTA}{cmyk}{0,1,0,0}
 \definecolor{YELLOW}{cmyk}{0,0,1,0}
 }

\@ifundefined{definecolor}{\usepackage{color}}{}
\usepackage{latexsym}\usepackage{bm}

\makeatother

\begin{document}

\title{AdS-Wave Solutions of $f$(Riemann) Theories}

\author{Metin G{\"u}rses}

\email{gurses@fen.bilkent.edu.tr}

\selectlanguage{english}%

\affiliation{{\small Department of Mathematics, Faculty of Sciences}\\
 {\small Bilkent University, 06800 Ankara, Turkey}}

\author{Sigbj{\o}rn Hervik}

\email{sigbjorn.hervik@uis.no}

\selectlanguage{english}%

\affiliation{Department of Mathematics and Natural Sciences, University of Stavanger,
N-4036 Stavanger, Norway}

\author{Tahsin \c{C}a\u{g}r\i{} \c{S}i\c{s}man}

\email{tahsin.c.sisman@gmail.com}

\selectlanguage{english}%

\affiliation{Department of Physics,\\
 Middle East Technical University, 06800 Ankara, Turkey}

\author{Bayram Tekin}

\email{btekin@metu.edu.tr}

\selectlanguage{english}%

\affiliation{Department of Physics,\\
 Middle East Technical University, 06800 Ankara, Turkey}
\begin{abstract}
We show that the recently found AdS-plane and AdS-spherical wave solutions
of quadratic curvature gravity also solve the most general higher
derivative theory in $D$-dimensions. More generally, we show that
the field equations of such theories reduce to an equation linear
in the Ricci tensor for Kerr-Schild spacetimes having type-N Weyl
and traceless Ricci tensors.
\end{abstract}
\maketitle
There is a vast literature on the exact solutions of four-dimensional
Einstein's gravity. But, as more powers of curvature are added, or
computed in a microscopic theory such as string theory, to get a better
UV behaved theory, the field equations become highly nontrivial and
so solutions are not easy to find. In fact, only a few classes of
solutions are known: For example, see \cite{Horowitz-Steif,Horowitz-Tseytlin,Coley,Hervik}
for solutions in low energy string theory. $\text{AdS}_{5}\times S^{5}$,
which played a major role in AdS/CFT, is also expected to be an exact
solution of string theory \cite{Banks-Green}. In this work, we present
new asymptotically AdS solutions, which are AdS-plane and AdS-spherical
waves, to $D$-dimensional generic gravity theories based on the Riemann
tensor and its arbitrary number of covariant derivatives which are
in some sense natural geometric extensions of Einstein's gravity.
Certain low energy string theory actions constitute a subclass of
this theory once all the non-gravitational fields are turned off \cite{Tseytlin}.
Asymptotically AdS solutions in higher derivative theories are relevant
in the context of generic gravity/gauge theory dualities and holography.
Here, we shall provide such solutions.

Using a theorem given in \cite{Hervik}, we first prove that any
spacetime with type-N Weyl and traceless Ricci tensors, where the
metric is the Kerr-Schild form, the field equations of the most
general higher derivative theory reduce to a linear equation for
the traceless Ricci tensor. These spacetimes have constant scalar
invariants. Furthermore, using the the type-N property of the
traceless Ricci tensor in these field equations, we obtain a
linear partial differential equation for the metric function $V$
of order $2N$ in the AdS background, where $N$ is related to the
number of covariant derivatives in the action of the theory. This
result implies that the AdS-wave metrics are universal in the
sense defined in \cite{Hervik}.

As a special case, the field equations of the theory which depends
on the contractions of the Riemann tensor but not on its derivatives,
$f\left(R_{\rho\sigma}^{\mu\nu}\right)$ theory, arealso highly cumbersome,
but using our general result for type-N spacetimes under certain assumptions
they reduce to those of the quadratic gravity. Then, taking the metric
to be in the Kundt subclass of type-N spacetimes we show that AdS-plane
wave \cite{Gullu-Gurses,Alishah} and AdS-spherical wave \cite{gurses1}
solutions of the quadratic gravity theory are also the solutions of
the $f\left(R_{\rho\sigma}^{\mu\nu}\right)$ theory. Log terms arising
in the solutions of the quadratic gravity exist also in some $f\left(R_{\rho\sigma}^{\mu\nu}\right)$
theories corresponding to the generalizations of critical gravity
\cite{LuPope,DeserLiu}. As an application of our result, we show
that any type-N Einstein space $\left(R_{\mu\nu}=\frac{R}{D}g_{\mu\nu}\right)$
solve the field equations of the Lanczos-Lovelock theory. In addition,
for a special choice of the parameters, any spacetime metric having
type-N Weyl and traceless Ricci tensors with constant Ricci scalar
solve the specific Lanczos-Lovelock theory.

To find exact asymptotically AdS solutions of a generic higher derivative
theory, we shall make use of general results of \cite{pravda1,pravda2,pravda3,pravda4,pravda5}
which utilize the boost weight formalism.

\paragraph*{Type-N Weyl and Traceless Ricci tensors}

The Weyl tensor of type-N spacetimes have been studied in detail in
\cite{pravda1,pravda2,pravda3,pravda4}. Let ${\bf \ell}$, ${\bf n}$
and ${\bf m}^{(i)}$, $(i=2,\cdots D-1)$ be a null tetrad frame with
\begin{eqnarray}
 &  & \ell_{\alpha}\,\ell^{\alpha}=n^{\alpha}\, n_{\alpha}=\ell^{\alpha}\, m_{\alpha}^{(i)}=n^{\alpha}\, m_{\alpha}^{(i)}=0,\nonumber \\
 &  & \ell^{\alpha}\, n_{\alpha}=1,~~m^{(i)\alpha}\, m_{\alpha}^{\left(j\right)}=\delta_{ij},
\end{eqnarray}
 where $\alpha,\beta,\cdots=0,1,2,\cdots D-1$ and $i,j=2,3,\cdots D-1$.
Then spacetime metric takes the form
\begin{equation}
g_{\mu\nu}=\ell_{\mu}\, n_{\nu}+\ell_{\nu}\, n_{\mu}+\delta_{ij}\, m_{\mu}^{(i)}\, m_{\nu}^{\left(j\right)}.\label{metric1}
\end{equation}
 The Weyl tensor of type-N spacetimes, expressed in the above frame,
where ${\bf \ell}$ is the Weyl aligned null direction takes the form
\begin{equation}
C_{\alpha\beta\gamma\delta}=4\Omega_{ij}^{\prime}\,\ell_{\{\alpha}m_{\beta}^{(i)}\,\ell_{\gamma}\, m_{\delta\}}^{(j)}.
\end{equation}
 which transforms under scale transformations with boost weight -2.

Type-N property alone is not sufficient to reduce the field equations
of higher derivative gravity theories to a solvable form, we make
a further assumption that the spacetime is radiating and the Ricci
scalar is constant. For radiating type-N spacetimes, the Ricci tensor
has the form
\begin{equation}
R_{\mu\nu}=\rho\ell_{\mu}\,\ell_{\nu}+\frac{R}{D}\, g_{\mu\nu},\label{eq:Null_Ricci}
\end{equation}
 where $\rho$ is a scalar function and $R$ is the Ricci scalar.
Taking $R$ to be a constant and using the Bianchi identity, one obtains
\begin{equation}
\nabla^{\mu}\,(\rho\,\ell_{\mu}\,\ell_{\nu})=0.\label{bianchi1}
\end{equation}
 Notice that the traceless part
\begin{equation}
S_{\mu\nu}=R_{\mu\nu}-\frac{R}{D}\, g_{\mu\nu}=\rho\,\ell_{\mu}\ell_{\nu}
\end{equation}
 of the Ricci tensor is of type-N. Following \cite{Hervik}, and paraphrasing
the statement given in this work we have the following result:

\vspace{0.4cm}

\noindent Theorem: \textit{Consider a Kundt spacetime for which the traceless Ricci and the Weyl tensors be
of type-N (\ref{eq:Null_Ricci}), and all scalar invariants be constant.
Then, any second rank symmetric tensor constructed by the Riemann
tensor and its covariant derivatives is a linear combination of $g_{\mu\nu}$,
$S_{\mu\nu}$, and higher orders of $S_{\mu\nu}$(such as, for example,
$\square^{n}\, S_{\mu\nu}$).}

\vspace{0.4cm}

\noindent The proof of the above statement depends heavily on the
boost-weight formalism \cite{Hervik},\cite{coley-hervik-pelavas-2006}, \cite{CHP}. In particular, a Kundt spacetime of Ricci and Weyl type N needs to be degenerate Kundt, which further implies, using the same arguments as in \cite{Hervik}, that the traceless part of any second rank symmetric tensor must also be of type N. Using the appendix of \cite{Hervik}, and the proof of theorem 2.7 in \cite{CHP}, one can now see that the theorem follows.

An immediate consequence of above theorem is as follows:

\vspace{0.4cm}

\noindent For the Kundt type of Kerr-Schild metrics where $S_{\mu\nu}=\rho\,\ell_{\mu}\ell_{\nu}$,
we get $\square^{n}\, S_{\mu\nu}=\ell_{\mu}\,\ell_{\nu}\mathcal{O}^{n}\,\rho$
for all $n=0,1,2,\cdots$ (we shall give the definition of the operator
$\mathcal{O}$ shortly), then the higher orders of $S_{\mu\nu}$ vanish
identically. This nice property of this Kerr-Schild metrics leads
to the following more general result. The field equations corresponding
to the most general action;
\begin{eqnarray}
 &  & I=\int\, d^{D}\, x\sqrt{-g}\, f(g^{\alpha\beta},R^{\mu}\,_{\nu\gamma\sigma},\nabla_{\rho}R^{\mu}\,_{\nu\gamma\sigma},\nonumber \\
 &  & \dots,\left(\nabla_{\rho_{1}}\nabla_{\rho_{2}}\dots\nabla_{\rho_{M}}\right)R^{\mu}\,_{\nu\gamma\sigma},\dots)\,,
\end{eqnarray}
 are of the form
\begin{equation}
E_{\mu\nu}=eg_{\mu\nu}+\sum_{n=0}^{N}a_{n}\square^{n}\, S_{\mu\nu}=0,\label{eq:EoM_generic}
\end{equation}
 where $e$ and $a_{n}$'s are all constants depending on the form
of the action. Hence we reduced the field equations of the most general
higher derivative theories to a form linear in the Ricci tensor. Furthermore,
the trace part of (\ref{eq:EoM_generic}) is
\begin{equation}
e=0,
\end{equation}
 which gives the relation between the cosmological constant and the
parameters of the theory, and the traceless part is
\begin{equation}
\sum_{n=0}^{N}a_{n}\square^{n}\left(\rho\ell_{\mu}\ell_{\nu}\right)=0.\label{eq:EoM_general}
\end{equation}
 This is still hard to solve: despite its appearance it is a nonlinear
equation. But, we know at least two solutions which are the AdS-plane
\textit{\emph{\cite{Gullu-Gurses,Alishah}}} and the AdS-spherical
waves \textit{\emph{\cite{gurses1}}}. These waves are of the Kerr-Schild
form, $g_{\mu\nu}=\bar{g}_{\mu\nu}+2V\ell_{\mu}\,\ell_{\nu}$, which
belong to the Kundt class of type-N spacetimes \cite{coley-hervik-pelavas-2006,ColeyHervik}.
Here, $\bar{g}_{\mu\nu}$ is the AdS background metric and the function
$V$ satisfies $\ell^{\mu}\,\partial_{\mu}V=0$. For these metrics,
$\rho$ has the form
\begin{equation}
\rho\equiv\mathcal{O}V=\left[\bar{\square}+2\xi_{\mu}\partial^{\mu}+\frac{1}{2}\xi_{\mu}\xi^{\mu}-2k^{2}\left(D-2\right)\right]V,\label{eq:rho}
\end{equation}
 where $\bar{\square}$ is the Laplace-Beltrami operator of the AdS
background and $\xi_{\mu}$ arises in $\nabla_{\mu}\ell_{\nu}=\ell_{(\mu}\xi_{\nu)}$.
Because of (\ref{eq:rho}), Eqn.~(\ref{eq:EoM_general}) reduces
to
\begin{equation}
\sum_{n=0}^{N}a_{n}\mathcal{O}^{n+1}V=0.
\end{equation}
 For $N>1$, this equation can be factorized as
\begin{equation}
\prod_{n=0}^{N}\left(\mathcal{O}+b_{n}\right)\,\mathcal{O}\, V=0,
\end{equation}
 where some $b_{n}$'s are real and some are complex constants in
general (complex $b_{n}$'s come in complex conjugate pairs). Then,
the most general solution is
\[
V=\Re\left(\sum_{i=0}^{N}V_{i}\right),
\]
where $\Re$ represents the real part and $V_{i}$'s solve $\left(\mathcal{O}+b_{i}\right)V_{i}=0$,
$i=0,1,2,\cdots,N$. Solutions of such linear partial differential
equations are given in \cite{Gullu-Gurses} and \cite{gurses1} in
the of AdS background. As a conclusion we can say that the AdS- wave
metrics found recently \cite{Gullu-Gurses} and \cite{gurses1} solve
the field equations of the most general higher derivative theories.
To give explicit exact solutions we have to know the constants $b_{i}$,
$i=0,1,2,\cdots$ in terms of the theory parameters. For this purpose
we shall consider below some special cases.

\vspace{0.4cm}

\paragraph*{Type-N Spacetimes in Quadratic Gravity --}

For type-N Weyl and type-N traceless Ricci tensor and (\ref{bianchi1})
the field equations of quadratic gravity \cite{DeserTekin}
\begin{align}
\frac{1}{\kappa}\left(R_{\mu\nu}-\frac{1}{2}g_{\mu\nu}R+\Lambda_{0}g_{\mu\nu}\right)\nonumber \\
+2\alpha R\left(R_{\mu\nu}-\frac{1}{4}g_{\mu\nu}R\right)+\left(2\alpha+\beta\right)\left(g_{\mu\nu}\square-\nabla_{\mu}\nabla_{\nu}\right)R\nonumber \\
+\beta\square\left(R_{\mu\nu}-\frac{1}{2}g_{\mu\nu}R\right)+2\beta\left(R_{\mu\sigma\nu\rho}-\frac{1}{4}g_{\mu\nu}R_{\sigma\rho}\right)R^{\sigma\rho}\nonumber \\
+2\gamma\biggl[RR_{\mu\nu}-2R_{\mu\sigma\nu\rho}R^{\sigma\rho}+R_{\mu\sigma\rho\tau}R_{\nu}^{\phantom{\nu}\sigma\rho\tau}-2R_{\mu\sigma}R_{\nu}^{\phantom{\nu}\sigma}\nonumber \\
-\frac{1}{4}g_{\mu\nu}\left(R_{\tau\lambda\sigma\rho}^{2}-4R_{\sigma\rho}^{2}+R^{2}\right)\biggr]=0 & ,\label{fieldequations}
\end{align}
 reduce to the following simplified equations \cite{pravda3,pravda4},
\begin{equation}
\left(\beta\square+c\right)\left(\rho\ell_{\mu}\ell_{\nu}\right)=0,
\end{equation}
 and a trace equation that gives a relation between the constant $R$
and the parameters of the theory. Here, $c$ is given as \cite{DeserTekin}
\begin{equation}
c\equiv\frac{1}{\kappa}+2R\alpha+\frac{2\left(D-2\right)}{D\left(D-1\right)}R\beta+\frac{2\left(D-3\right)\left(D-4\right)}{D\left(D-1\right)}R\gamma.\label{eq:c}
\end{equation}
 Exact solutions, the AdS-wave metrics, of this have been reported
recently \textcolor{blue}{\cite{Alishah,Gullu-Gurses}} and \cite{gurses1}.

\vspace{0.4cm}

\paragraph*{$f\left(R_{\rho\sigma}^{\mu\nu}\right)$theory--}

A case which is more general then the quadratic gravity is the $f$
(Riemann) theory where the action is given by
\begin{equation}
I=\int\, d^{D}\, x\,\sqrt{-g}\, f\left(R_{\rho\sigma}^{\mu\nu}\right),
\end{equation}
 For this case $N=1$ and hence (\ref{eq:EoM_general}) reduces to
\begin{equation}
(a\square+b)(\rho\ell_{\mu}\ell_{\nu})=0,\label{eq:EoM}
\end{equation}
 where $a$ and $b$ are constants. For AdS-wave metrics in the Kerr-Schild
form, $g_{\mu\nu}=\bar{g}_{\mu\nu}+2V\ell_{\mu}\,\ell_{\nu}$, the
field equations reduce to a fourth order linear partial differential
equation with constant coefficients of the form
\begin{equation}
\left(\bar{\square}-\frac{2R}{D\left(D-1\right)}-M^{2}\right)\left(\bar{\square}-\frac{2R}{D\left(D-1\right)}\right)\left(V\ell_{\mu}\ell_{\nu}\right)=0,\label{eq:AdS-wave_eom}
\end{equation}
 where $\bar{\square}$ is the Laplace-Beltrami operator of the AdS
background and
\begin{equation}
M^{2}=-\frac{b}{a}-\frac{2R}{D\left(D-1\right)},
\end{equation}
 which corresponds to the mass of the spin-2 excitation in the linearized
version of $f$(Riemann) theory about AdS \cite{Senturk}. For AdS-plane
wave, $\bar{g}_{\mu\nu}=-\frac{R}{D\left(D-1\right)z^{2}}\eta_{\mu\nu}$
where $z>0$ is one of the spatial coordinates, then $\ell_{\mu}=\left(1,1,0,\dots,0\right)$,
$\xi_{\mu}=\frac{2}{z}\,\delta_{\mu}^{z}$ and the most general solution
of (\ref{eq:AdS-wave_eom}) was given in \cite{Gullu-Gurses}. In
this case $\ell_{\mu}$ is proportional to the null Killing vector
$\zeta_{\mu}$, i.e., $\zeta_{\mu}=\frac{1}{z}\,\ell_{\mu}$. For
AdS-spherical wave, $\ell_{\mu}=\left(1,\frac{x^{i}}{r}\right)$,
$\xi_{\mu}=-\frac{1}{r}\,\ell_{\mu}+\frac{2}{r}\,\delta^{t}\,_{\mu}+\frac{2}{z}\,\delta^{z}\,_{\mu}$
where $r^{2}=\sum_{i=1}^{D-1}\left(x^{i}\right)^{2}$ and $x^{D-1}=z$,
and the most general solution of (\ref{eq:AdS-wave_eom}) was given
in \cite{gurses1}. In this case there exists no null Killing vector
fields. For $M^{2}\ne0$, $V$ decays sufficiently fast for both AdS-wave
solutions and they are asymptotically AdS. When $M^{2}=0$, there
are logarithmic solutions which spoil the asymptotically AdS structure
in both cases. Let us give two concrete $f\left(R_{\rho\sigma}^{\mu\nu}\right)$
theories as examples.

\paragraph*{Cubic gravity generated by string theory--}

In \cite{Tseytlin}, the bosonic string has the following effective
Lagrangian density at $O\left[\left(\alpha^{\prime}\right)^{2}\right]$
\begin{align}
f_{\text{eff}}= & R+\frac{\alpha^{\prime}}{4}\left(R_{\alpha\beta}^{\mu\nu}R_{\mu\nu}^{\alpha\beta}-4R_{\nu}^{\mu}R_{\mu}^{\nu}+R^{2}\right)\label{eq:Effective_string}\\
 & +\frac{\left(\alpha^{\prime}\right)^{2}}{24}\left(-2R^{\mu\alpha\nu\beta}R_{\mu\phantom{\lambda}\nu}^{\phantom{\mu}\lambda\phantom{\nu}\gamma}R_{\alpha\gamma\beta\lambda}+R_{\alpha\beta}^{\mu\nu}R_{\mu\nu}^{\gamma\lambda}R_{\gamma\lambda}^{\alpha\beta}\right),\nonumber
\end{align}
 where $\alpha^{\prime}$ is the usual inverse string tension. Using
the results of \cite{Sisman}, the field equations of (\ref{eq:Effective_string})
for the AdS-plane and AdS-spherical wave metrics reduce to (\ref{eq:AdS-wave_eom})
with $a$ and $M^{2}$ given as
\begin{equation}
a=\frac{7\alpha^{\prime2}R}{4D\left(D-1\right)},
\end{equation}
\begin{align}
M^{2}= & -\frac{4D\left(D-1\right)}{7\alpha^{\prime2}R}-\frac{2\left(D-3\right)\left(D-4\right)}{7\alpha^{\prime}}\nonumber \\
 & +\frac{9D-29}{7D\left(D-1\right)}R.
\end{align}
 Therefore, the solutions quoted above are the solutions of (\ref{eq:Effective_string})
with this $M^{2}$.

\paragraph*{Lanczos-Lovelock theory--}

Another special case of the $f$(Riemann) theory is the Lanczos-Lovelock
theory given with the Lagrangian density
\begin{equation}
f_{\text{L-L}}=\sum_{n=0}^{\left[\frac{D}{2}\right]}a_{n}\delta_{\nu_{1}\dots\nu_{2n}}^{\mu_{1}\dots\mu_{2n}}\prod_{p=1}^{n}R_{\mu_{2p-1}\mu_{2p}}^{\nu_{2p-1}\nu_{2p}},\label{eq:Lovelock}
\end{equation}
 where $a_{n}$'s are dimensionful constants, $\left[\frac{D}{2}\right]$
corresponds to the integer part of $\frac{D}{2}$ and $\delta_{\nu_{1}\dots\nu_{2n}}^{\mu_{1}\dots\mu_{2n}}$
is the generalized Kronecker delta. In this case, since the constant
$a$ in (\ref{eq:EoM}) vanishes identically, the field equations
reduce to
\begin{equation}
b\rho=0,\label{eq:LL-EoM}
\end{equation}
 where $b$ is calculated in \cite{Gullu} as
\begin{equation}
b=2\left(D-3\right)!\sum_{n=0}^{\left[\frac{D}{2}\right]}a_{n}\frac{n\left(D-2n\right)}{\left(D-2n\right)!}\left[\frac{4\Lambda}{\left(D-1\right)\left(D-2\right)}\right]^{n-1}.
\end{equation}
 Equation (\ref{eq:LL-EoM}) gives two subclasses. First class corresponds
to $\rho=0$ and the solution is type-N Einstein space. The second
class corresponds to $b=0$ which gives a relation between the parameters
of the theory. In this subclass, any type-N radiating metric with
constant Ricci scalar is an exact solution of the theory. AdS-wave
metrics are exact solutions of the Lancsoz-Lovelock theory. We note
that the Lanczos-Lovelock theory is free of logarithmic solutions.

\vspace{0.4cm}

\paragraph*{Conclusion--}

Type-N radiating spacetimes simplify the field equations of higher
derivative theories. This simplification rests on the result: the
second rank symmetric  tensors constructed by contractions of
type-N Weyl, and traceless Ricci tensors and their covariant
derivatives reduce to a simpler form in general. To find explicit
solutions, we showed that one has to consider a subclass that is
the Kundt spacetime. Hence, Kerr-Schild metrics reported recently
as the solutions of the quadratic gravity, the AdS-plane
\cite{Gullu-Gurses} and AdS-spherical wave metrics \cite{gurses1},
which belong to the Kundt subclass of type-N spacetimes solve the
field equations of any higher curvature gravity theory exactly. We
gave a cubic theory coming from the string theory and the
Lanczos-Lovelock theory as explicit examples.

A more detailed version of this work will be communicated elsewhere.

\vspace{0.4cm}
 This work is partially supported by the Scientific and Technological
Research Council of Turkey (TUBITAK).


\begin{thebibliography}{10}
\bibitem{Horowitz-Steif} G.~T.~Horowitz and A.~R.~Steif, Phys.\ Rev.\ Lett.~\textbf{64},
260 (1990).

\bibitem{Horowitz-Tseytlin} G.~T.~Horowitz and A.~A.~Tseytlin,
Phys.\ Rev.\ D \textbf{51}, 2896 (1995).

\bibitem{Coley} A.~A.~Coley, Phys.\ Rev.\ Lett.\ \textbf{89},
281601 (2002).

\bibitem{Hervik} A.~A.~Coley, G.~W.~Gibbons, S.~Hervik and C.~N.~Pope,
Class.\ Quant.\ Grav.\ \textbf{25}, 145017 (2008).

\bibitem{Banks-Green} T.~Banks and M.~B.~Green, JHEP \textbf{9805},
002 (1998).

\bibitem{Tseytlin} R.~R.~Metsaev and A.~A.~Tseytlin, Phys.\ Lett.\ B
\textbf{185}, 52 (1987).

\bibitem{Gullu-Gurses} I.~Gullu, M.~Gurses, T.~C.~Sisman and
B.~Tekin, Phys.\ Rev.\ D \textbf{83}, 084015 (2011).

\bibitem{Alishah} M.~Alishahiha and R.~Fareghbal, Phys.\ Rev.\ D
\textbf{83}, 084052 (2011).

\bibitem{gurses1} M.~Gurses, T.~C.~Sisman and B.~Tekin, Phys.\ Rev.\ D
\textbf{86}, 024009 (2012).

\bibitem{LuPope} H.~Lu and C.~N.~Pope, Phys.\ Rev.\ Lett.\ \textbf{106},
181302 (2011).

\bibitem{DeserLiu} S.~Deser, H.~Liu, H.~Lu, C.~N.~Pope, T.~C.~Sisman
and B.~Tekin, Phys. Rev. \textbf{D83}, 061502 (2011).

\bibitem{pravda1} M. Ortaggio, V. Pravda and A. Pravdova, Class.\ Quantum
Grav.\textbf{\ 30} 013001, (2013).

\bibitem{pravda2} T.~Malek and V.~Pravda, Class.\ Quantum Grav.\textbf{\ 28},125011,
(2011).

\bibitem{pravda3} T.~Malek and V.~Pravda, Phys.\ Rev.\ D \textbf{84},
024047 (2011).

\bibitem{pravda4} M.~Ortaggio, V.~Pravda and A.~Pravdova, Phys.\ Rev.\ D
\textbf{82}, 064043 (2010).

\bibitem{pravda5} A.~Coley, R.~Milson, V.~Pravda and A.~Pravdova,
Class. Quantum Grav.\textbf{\ 21} (2004) 5519.

\bibitem{DeserTekin} S.~Deser and B.~Tekin, Phys.\ Rev.\ Lett.\textbf{\ 89},
101101 (2002); Phys.\ Rev.\ \textbf{D67}, 084009 (2003).

\bibitem{CHP}
  A.~Coley, S.~Hervik and N.~Pelavas,
  Class. Quant. Grav.  {\bf 27}, 102001 (2010).

\bibitem{coley-hervik-pelavas-2006} A.~Coley, S.~Hervik and N.~Pelavas,
Class.\ Quant.\ Grav.\ \textbf{23}, 3053-3074 (2006).

\bibitem{ColeyHervik} A.~Coley, S.~Hervik, G.~O.~Papadopoulos
and N.~Pelavas, Class.\ Quant.\ Grav.\ \textbf{26}, 105016 (2009).

\bibitem{Senturk} C.~Senturk, T.~C.~Sisman and B.~Tekin, Phys.\ Rev.\ D
\textbf{86}, 124030 (2012).

\bibitem{Sisman}T.~C.~Sisman, I.~Gullu and B.~Tekin, Class.\ Quant.\ Grav.\ \textbf{28},
195004 (2011).

\bibitem{Gullu} T.~C.~Sisman, I.~Gullu and B.~Tekin, Phys.\ Rev.\ D
\textbf{86}, 044041 (2012).\end{thebibliography}
\end{document}